# What is expected for China's SARS-CoV-2 epidemic?

**Carlos Hernandez-Suarez** [1]**, Efrén Murillo-Zamora** [2,*]**,**

[1]*Instituto de Ciencias Tecnología e Innovación, Universidad Francisco Gavidia, El Progreso St., No. 2748, Colonia Flor Blanca, San Salvador, El Salvador.*

[2]*Departamento de Epidemiología, Unidad de Medicina Familiar No. 19, IMSS, Av. Javier Mina 301, 28000, Colima, Colima. MEXICO.*

Correspondence*:
Corresponding Author
efren.murilloza@imss.gob.mx

**ABSTRACT**

Recently, China announced that its "zero-covid' policy would end, which will bring serious challenges to the country's health system. In here we provide simple calculations that allows us to provide an estimate of what is expected as an outcome in terms of fatalities, using the fact that it is a highly contagious disease that will expose most of a highly vaccinated population to the virus. We use recent findings regarding the amount of reduction in the risk of severe outcome achieved by vaccination and arrive to an estimate of 1.1 m deaths, 60% of these are males. In our model, 84% percent of deaths occur in individuals with age 55 years or older. In a scenario in which this protection is completely lost due to waning and the infection fatality rate of the prevalent strain reaches similar levels to the observed in the beginning of the epidemic, the death toll could reach 2.4 m, 93% in 55 years or older.

## 1  INTRODUCTION

China was the first country to report SARS-CoV-2 cases in December 2019 (Pekar et al., 2021) and to implement a "zero-covid' policy, consisting in mass testing, centralized quarantine, and lockdowns. On December 7[th], 2022, the government announced that this policy is ending (Wall Street Journal, 2022) which may bring great challenges to the country (Chen and Chen, 2022). Several studies have shown that the basic reproductive number $R_0$, the potential of disease transmission in the absence of any control or mitigation measures, may be as high as 5 (Sanche et al., 2020; D'Arienzo and Coniglio, 2020; Liu et al., 2020; Alimohamadi et al., 2020), which suggests that more than 98% of the Chinese population will be exposed to the virus. Nevertheless, 92.61% of the population has received at least one dose of vaccine (Johns Hopkins University, 2022a), which will affect the number of expected fatalities. To date (2022/12/24), there has been reported 16,628 deaths (Johns Hopkins University, 2022a) out of a population of about 1.4 billion, which is a reflect of the stringent "zero-covid" policy that has been implemented.

Early in the pandemics, we made forecasts of the death toll in a Za'atari refugee camp in Jordan (Hernandez-Suarez et al., 2020) by analyzing the fatality rate of SARS-CoV-2 at different ages in Mexico and projecting these in the refugee camp by using its demography. Nevertheless, this was before vaccines were available. Now, there is ample evidence of a waning vaccine effect (Goldberg et al. 2021; Housset et al. 2022; Levine-Tiefenbrun et al., 2022; Coppeta et al., 2022; Berar-Yanay et al., 2021; Kolaric et al., 2021; Hernandez-Suarez and Murillo-Zamora, 2022) and there is evidence that vaccines increase in general $Age_{50}$ (the age at which there is a 50% probability of severe disease if infected) in about 15 years on average for all vaccines (Hernandez-Suarez and Murillo-Zamora, 2022). In what follows we take all these factors into consideration to build a projection on the expected number of fatalities in China.

There are several factors that we cannot consider here that are indeed relevant, for instance, the availability of medical resources, including doctors, nurses, hospitals, ICU beds, new medications and the distribution and number of known comorbidities in the population. Nevertheless, even a large availability of these resources could be overwhelmed if the epidemic in China grows exponentially. Also, there is little information on the lethality of current and possible new strains, therefore, our analysis relies on the observed average lethality, the expected protection of vaccines and the available knowledge on demographics.

## 2 METHODOLOGY

In Hernandez-Suarez et al. (2020), we reported the observed lethality per group age, shown in Table 1 using data from a Mexican unvaccinated population with 10,330 SARS-CoV-2 infections. Recall the numbers in Table 1 are the proportion of individuals of each group age and gender that die from the disease, if infected. As previously mentioned, given the high $R_0$- value we assume that most of the Chinese population will be infected.

**Table 1.** Lethality of SARS-CoV-2 in an unvaccinated Mexican population

| Age | Males | Females |
| --- | --- | --- |
| 0-9 | 0.000028 | 0.000008 |
| 10-19 | 0.000017 | 0.000015 |
| 20-29 | 0.000155 | 0.000145 |
| 30-39 | 0.000811 | 0.000381 |
| 40-49 | 0.001998 | 0.001258 |
| 50-59 | 0.003543 | 0.002012 |
| 60 + | 0.005871 | 0.004596 |

Table 2 contains the Chinese demographics by group age (National Bureau of Statistics of China, 2021). Since Sinovac, the prevalent vaccine was found to increase $AGE_{50}$ in 16 years on average, what is needed is to subtract 16 years to each individual and rebuild the age categories. Since available intervals are 5 years wide, the closest multiple of 5 to 16 years is 15, and so we subtract 15 years to each boundary. All negative boundaries were converted to zero. Then, we concentrated the data in seven age groups to match Table 1. The result is the adjusted Chinese population shown in Table 3.

## 3  RESULTS

The element-by-element product of Tables 1 and 3 yields Table 4, containing expected deaths by group age and gender, for an expected total of 1.1 m deaths. About 60% of these are males and about the 84% of deaths would occur in 55 years or older.

## 4  DISCUSSION

Initially, this methodology was developed as a preparation for the expected outcome in refugee camps, whose crowded conditions would guarantee a fast and broad dissemination of the virus. Nevertheless, as mentioned before, the basic reproductive number of the disease is high enough to guarantee that eventually, a broad dissemination of SARS-CoV-2 would occur in any population. Our model does not allow to provide any time frame, only death toll. In October 2020 we applied this methodology to the Za'atari refugee camp in Jordan and projected 92 deaths out of 83,465 refugees, whereas for Kutupalong-Balukhali Site in Bangladesh, with 600 k refugees, we projected 565 deaths. This implied a 0.1% death toll. So far, the death toll in Za'atari camp is 48 (UNHCR 2022) and in Kutupalong-Balukhali camp is 42 (WHO 2022). The large differences are due to our model is very sensitive to the demographics of the population, for instance a 60+ years infected individual is 7 times more likely to die than 30 years old, and the demographics of a refugee population is difficult to assess, and thus a lot of assumptions had to be made regarding demographics of the camps. Nevertheless, our main finding was that our model predicted a fifth of deaths of other studies (Truelove, 2020), and contributed to eradicate the idea that poor nourishment, lack of medical resources, poverty, crowdedness and other diseases, characteristics of refugee camps, would rampage the population.

When we applied our methodology to Cuba, a country with similar characteristics than China in vaccine distribution (94.63% vaccinated) which also developed and deployed a national vaccine, and whose demography is available (United Nations, 2021), we calculated an expected death toll of 10,588 (see detailed calculations in Supplementary material). To date, the reported number of deaths in Cuba is 8,530, and no deaths have been reported in the last six months (Johns Hopkins University, 2022b). Our estimate is not far from the observed especially considering the possibility of under-reporting for many natural reasons.

For China, The Economist (2022) predicted around 1.5 million deaths using a modification of the SEIR model of Cai et al. (2022) and Airfinity, a data firm, estimated between 1.3 and 2.1 million deaths if the "zero-covid" policy were to be lifted (Airfinity, 2022).

Our calculations assume a protective effect of the vaccine according to what it has been reported in Hernandez-Suarez and Murillo-Zamora, (2022). Nevertheless, in a scenario in which this protection is completely lost due to waning, and the infection fatality rate of the prevalent strain reaches values around those observed in the beginning of the epidemic, the death toll could reach 2.4 m. 93% in 55 years or older.

**Table 2.** Demographics of Chinese population per group age.

| Age | Males | Females | Total |
|---|---|---|---|
| 0-4 | 40,969,331 | 36,914,557 | 77,883,888 |
| 5-9 | 48,017,458 | 42,226,598 | 90,244,056 |
| 10-14 | 45,606,790 | 39,649,204 | 85,255,994 |
| 15-19 | 39,053,343 | 33,630,797 | 72,684,140 |
| 20-24 | 39,675,995 | 35,265,680 | 74,941,675 |
| 25-29 | 48,162,270 | 43,685,062 | 91,847,332 |
| 30-34 | 63,871,808 | 60,273,382 | 124,145,190 |
| 35-39 | 50,932,037 | 48,080,895 | 99,012,932 |
| 40-44 | 47,632,694 | 45,322,636 | 92,955,330 |
| 45-49 | 58,191,686 | 56,033,201 | 114,224,887 |
| 50-54 | 61,105,470 | 60,058,826 | 121,164,296 |
| 55-59 | 50,816,026 | 50,584,760 | 101,400,786 |
| 60-64 | 36,871,125 | 36,511,813 | 73,382,938 |
| 65-69 | 36,337,923 | 37,667,637 | 74,005,560 |
| 70-74 | 24,162,733 | 25,427,303 | 49,590,036 |
| 75-79 | 14,752,433 | 16,486,416 | 31,238,849 |
| 80+ | 15,257,272 | 20,543,563 | 35,800,835 |
|  | 721,416,394 | 688,362,330 | 1,409,778,724 |

**Table 3.** Adjusted Demographics of Chinese population per group age (after reducing 15 years to boundaries to reflect the vaccine effect).

| Age | Males | Females | Total |
|---|---|---|---|
| 0-9 | 213,322,917 | 187,686,836 | 401,009,753 |
| 10-19 | 112,034,078 | 103,958,444 | 215,992,522 |
| 20-29 | 98,564,731 | 93,403,531 | 191,968,262 |
| 30-39 | 119,297,156 | 116,092,027 | 235,389,183 |
| 40-49 | 87,687,151 | 87,096,573 | 174,783,724 |
| 50-59 | 60,500,656 | 63,094,940 | 123,595,596 |
| 60 + | 30,009,705 | 37,029,979 | 67,039,684 |
| Total | 721,416,394 | 688,362,330 | 1,409,778,724 |

**Table 4.** Expected deaths per group age, China (15 years have been added to boundaries)

| Age   | Males   | Females | Total     |
|-------|---------|---------|-----------|
| 0-24  | 5,973   | 1,501   | 7,475     |
| 25-34 | 1,905   | 1,559   | 3,464     |
| 35-44 | 15,278  | 13,544  | 28,821    |
| 45-54 | 96,750  | 44,231  | 140,981   |
| 55-64 | 175,199 | 109,567 | 284,766   |
| 65-74 | 214,354 | 126,947 | 341,301   |
| 75 +  | 176,187 | 170,190 | 346,377   |
| Total | 685,645 | 467,540 | 1,153,185 |

# Supplementary Material

**Table S1.** Lethality of SARS-CoV-2 in an unvaccinated Mexican population.

| Age   | Males    | Females  |
|-------|----------|----------|
| 0-9   | 0.000028 | 0.000008 |
| 10-19 | 0.000017 | 0.000015 |
| 20-29 | 0.000155 | 0.000145 |
| 30-39 | 0.000811 | 0.000381 |
| 40-49 | 0.001998 | 0.001258 |
| 50-59 | 0.003543 | 0.002012 |
| 60 +  | 0.005871 | 0.004596 |

**Table S2.** Demographics of Cuban population per group age.

| Age   | Males     | Females   | Total      |
|-------|-----------|-----------|------------|
| 0-4   | 284,758   | 265,909   | 550,666    |
| 5-9   | 317,987   | 297,439   | 615,426    |
| 10-14 | 302,634   | 285,574   | 588,208    |
| 15-19 | 321,819   | 304,354   | 626,173    |
| 20-24 | 358,731   | 338,129   | 696,860    |
| 25-29 | 366,550   | 343,788   | 710,338    |
| 30-34 | 408,870   | 386,649   | 795,519    |
| 35-39 | 353,346   | 340,870   | 694,216    |
| 40-44 | 322,242   | 316,024   | 638,266    |
| 45-49 | 450,573   | 454,723   | 905,296    |
| 50-54 | 480,598   | 498,770   | 979,368    |
| 55-59 | 462,561   | 492,306   | 954,867    |
| 60-64 | 304,748   | 333,958   | 638,706    |
| 65-69 | 256,611   | 287,734   | 544,345    |
| 70-74 | 210,979   | 243,047   | 454,026    |
| 75-79 | 153,365   | 186,785   | 340,150    |
| 80-84 | 97,694    | 124,981   | 222,675    |
| 85-89 | 48,696    | 66,613    | 115,309    |
| 90-94 | 20,235    | 30,807    | 51,042     |
| 95-99 | 7,092     | 10,628    | 17,720     |
| 100 + | 3,487     | 4,731     | 8,218      |
| Total | 5,533,576 | 5,613,819 | 11,147,394 |

**Table S3.** Adjusted Demographics of Cuban population per group age (after reducing 15 years to boundaries to reflect the vaccine effect).

| Age | Males | Females | Total |
|---|---|---|---|
| 0-9 | 1,585,929 | 1,491,405 | 3,077,333 |
| 10-19 | 775,420 | 730,437 | 1,505,857 |
| 20-29 | 675,588 | 656,894 | 1,332,482 |
| 30-39 | 931,171 | 953,493 | 1,884,664 |
| 40-49 | 767,309 | 826,264 | 1,593,573 |
| 50-59 | 467,590 | 530,781 | 998,371 |
| 60 + | 330,569 | 424,545 | 755,114 |
| Total | 5,533,576 | 5,613,819 | 11,147,394 |

**Table S4.** Expected deaths per group age, Cuba (obtained multiplying element-by-element Tables S1 and S3 and adding 15 years to boundaries).

| Age | Males | Females | Total |
|---|---|---|---|
| 0-24 | 44 | 12 | 56 |
| 25-34 | 13 | 11 | 24 |
| 35-44 | 105 | 95 | 200 |
| 45-54 | 755 | 363 | 1,118 |
| 55-64 | 1,533 | 1,039 | 2,573 |
| 65-74 | 1,657 | 1,068 | 2,725 |
| 75 + | 1,941 | 1,951 | 3,892 |
| Total | 6,048 | 4,540 | 10,588 |